# Millimeter-wave study of London penetration depth temperature dependence in Ba(Fe$_{0.926}$Co$_{0.074}$)$_2$As$_2$ single crystal

A.A. Barannik[1], N.T. Cherpak[1], N. Ni[2], M.A. Tanatar[3], S.A. Vitusevich[4],

V.N. Skresanov[1], P.C. Canfield[2,3], R. Prozorov[2,3], V.V. Glamazdin[1], K.I. Torokhtii[5]

[1] *A. Usikov Institute of Radiophysics and Electronics of the National Academy of Sciences of Ukraine*
*12 Acad. Proskura str., Kharkiv 61085, Ukraine,*
e-mail: cherpak@ire.kharkov.ua

[2] *Department of Physics and Astronomy, Iowa State University, Ames, IA 50011, USA*

[3] *Ames Laboratory USDOE, Ames, IA 50011, USA*

[4] *Institute of Bio- and Nanostructures, Forschungszentrum Juelich,*
*1 Leo-Brandt Strasse, 52425 Juelich, Germany*

[5] *Physical Engineering Department, National Technical Institute-KhPI,*
*21 Frunze str., 61002 Kharkiv, Ukraine*

**In-plane surface Ka-band microwave impedance of optimally doped single crystals of the Fe-based superconductor Ba(Fe$_{0.926}$Co$_{0.074}$)$_2$As$_2$ (T$_c$= 22.8K) was measured. Sensitive sapphire disk quasi-optical resonator with high-Tc cuprate conducting endplates was developed specially for Fe-pnictide superconductors. It allowed finding temperature variation of London penetration depth in a form of power law, namely $\Delta\lambda(T) \sim T^n$ with $n$ = 2.8 from low temperatures up to at least 0.6$T_c$ consisted with radio-frequency measurements. This exponent points towards nodeless state with pairbreaking scattering, which can support one of the extended s-pairing symmetries. The dependence $\lambda(T)$ at low temperatures is well described by one superconducting small-gap ($\Delta \cong 0.75$ in $kT_c$ units, where $k$ is Boltzman coefficient) exponential dependence.**

The discovery of superconductivity in Fe-based compounds [1] stimulated tremendous efforts to establish their physical properties. Fe-based materials are similar to cuprate high-Tc superconductors, for which the mechanism of superconductivity still remains a mystery, although the symmetry of their energy gap was identified [2]. In Fe-based compounds the gap structure, particularly the presence or absence of nodes, is still a controversial issue despite the large number of the published works (see e.g. [3-7] and references therein). So far, full spectrum of possible gap structures, nodeless and nodal, single and multivalued, constant and sign – changing, has been suggested for Fe-based superconductors [2-9].

London penetration depth is a very informative directly measurable quantity. It connects temperature-dependent superfluid density with the superconducting gap and electronic band structure [8]. One of effective ways to determine temperature dependence of London penetration depth $\lambda$ is to measure surface impedance depending on temperature by using microwave resonators [2]. Since $\lambda$ is much larger than crystal lattice spacing, microwave measurements probe bulk properties, as opposite to surface techniques, such as ARPES and scanning tunnel spectroscopy. Combining a temperature dependence of the sample microwave surface impedance with a value of the known $\lambda$ measured by means of other technique we obtain possibility to determine absolute values of complex conductivity and its temperature dependence.

There has been a large number of studies of the penetration depth in FeAs-based superconductors by using tunnel-diode resonator technique at radio frequencies [10-12], by SQUID susceptometer and magnetic-force microscopy at low frequencies [13] as well as by static muon-spin rotation experiments [11,14]. At the very high frequency end, THz and optical spectroscopy has been reported [15-18]. However, we are aware of only two reports as for the microwave response of Fe-based superconductors, namely, in electron-doped PrFeAsO$_{1-y}$ ($y \cong 0.1$) [3] and hole-doped Ba$_{1-x}$K$_x$Fe$_2$As$_2$ ($x \cong 0.55$) [4] crystals. Both works have concluded multi-gap superconductivity without nodes and Ref.[4] emphasized possible influence of the impurity scattering effects.

Early radio-frequency measurements of $\lambda$ have revealed power-law behavior, $\Delta\lambda \sim T^\beta$ ($\beta \cong 2$) [7,18], which was interpreted to be either due to point nodes in clean case or due to scattering. Later systematic studies, however, showed that in order to understand variety of experimental results [19,20], one has to conclude that in-plane superconductivity in optimally doped samples is fully gapped, but shows definite features of so-called s$_{+/-}$ pairing.

Important advantage of microwave impedance measurements are high accuracy under condition of high Q-factor of microwave resonator and possibility to determine physical properties of superconductor electron system. In this Letter, we focus on the Ba(Fe$_{1-x}$Co$_x$)$_2$As$_2$ member of the 122 Fe-pnictide family, for which high quality single crystals are available. We report measurements of the in-plane temperature-dependent microwave surface reactance, $X_s$, as a part of impedance $Z_s = R_s + iX_s$ in optimally doped single crystals Ba(Fe$_{0.926}$Co$_{0.074}$)$_2$As$_2$ with critical temperature, $T_c = 22.8$ K. The more detailed results concerning surface impedance properties of this superconductor will be presented elsewhere. The crystals were grown from FeAs:CoAs flux, as described in [21]. Microwave measurements were performed in a Ka-band (35 – 40 GHz range) using sapphire disk quasi-optical resonator excited at whispering gallery modes (WGM). The known resonator with conducting endplates (CEP) [22] was modified into the disk resonator with a radial slot (see inset in Fig.1). This resonator geometry was applied for the first time in [23] by using CEPs made of YBa$_2$Cu$_3$O$_{7-\delta}$ films of $T_c \cong 90$K. It was developed specifically for $Z_s$ precision measurements of small-sized superconductors with $T_c < 90$K, i.e. $T_c$ of YBa$_2$Cu$_3$O$_{7-\delta}$.

All of peculiarities of the resonator allow one to obtain high Q-factor, namely, $Q \cong 10^5$ in temperature interval from LHe temperatures up to about 30K. We also developed a novel technique for processing the frequency response of the resonators with partial removal of mode degeneration [22] and perturbed resonance Lorenz line, which allowed us precise determination of the resonance frequency and the Q-factor and thus accurate finding $Z_s$.

The results of resonant frequency $f(T)$ measurement of the resonator with and without the studied crystal are shown in Fig.1.

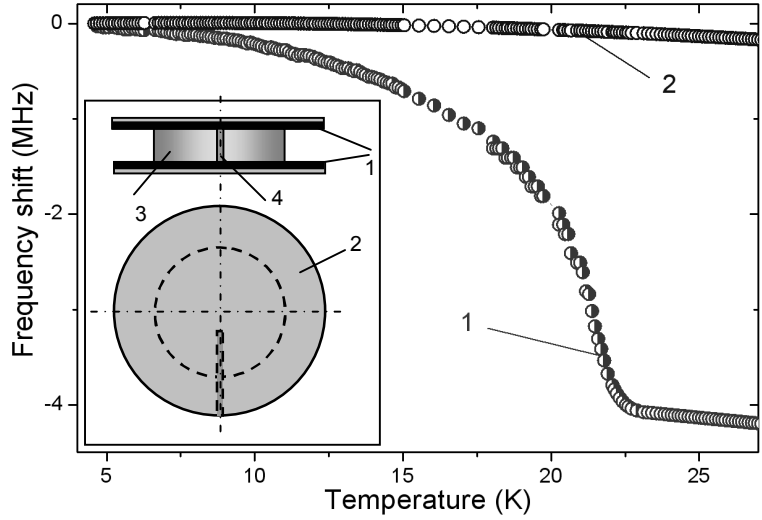

*Fig. 1.* The resonant frequency shift of the resonator with (curve 1) and without (curve 2) single crystal $Ba(Fe_{1-x}Co_x)_2As_2$ sample depending on temperature. Inset shows the slotted sapphire disk resonator with a single crystal $Ba(Fe_{1-x}Co_x)_2As_2$ in a slot. The superconducting films (1) are sputtered on the single crystal sapphire substrates (2), a sapphire disk (3) with a single crystal $Ba(Fe_{1-x}Co_x)2As_2$ (4) in a radial slot is sandwiched between superconducting $YBa_2Cu_3O_{7-\delta}$ endplates (1).

To obtain $X_s(T)$ from measured $f(T)$ we use the well-known expressions (see e.g. [22,24,25]). One can obtain expression for temperature variation of the surface reactance $\Delta X_s(T)$ through the temperature changing the resonator frequency $\Delta\omega(T)$

$$A_s \Delta X_s(T) = -2\Delta\omega(T)/\omega(T), \qquad (1)$$

where $\omega = 2\pi f$, $A_s$ is the inclusion coefficient of the sample under test. It depends on geometry and dimensions of the sample and field structure (mode) in the resonator. In a given work $A_s$ was evaluated by simulation of the resonator using Microwave Studio CST. We obtain $A_s = 2.83 \cdot 10^{-4}$ mOhm$^{-1}$ at interaction of $HE_{nm0}$-mode with a sample of $2.50 \times 3.50 \times 0.10$ mm$^3$ dimensions.

Evidently, in a case of WGM slotted resonator (see inset in Fig.1), analogously to other resonator techniques, the most appropriate approach can be one, at which variation $X_s(T)$ is determined [25] as

$$\Delta X_s(T, T_{ref}) = X_s(T, T_{ref}) - X_s(T_{ref}), \qquad (2)$$

where $T_{ref}$ is a certain reference temperature. Because $X_s(T) = \omega(T)\mu_0\lambda(T)$ at $T<<T_c$, we can write

$$\Delta X_s(T, T_{ref}) = \omega(T)\mu_0\Delta\lambda(T, T_{ref}), \qquad (3)$$

where $\Delta\lambda(T, T_{ref}) = \lambda(T) - \lambda(T_{ref})$. From (1) and (3) $\Delta\lambda(T, T_{ref})$ can be expressed as

$$\Delta\lambda(T, T_{ref}) = -\frac{2\Delta\omega(T, T_{ref})}{A_s \omega^2(T)\mu_0}, \qquad (4)$$

where $\Delta\omega(T, T_{ref}) = \omega(T) - \omega(T_{ref})$.

The experimental temperature law $\Delta\lambda(T, T_{ref})$ allows one to extrapolate it to $T \to 0$ and, knowing $\lambda(0)$ from other measurements, to determine $\lambda(T)$.

It is worthy to note that in $\Delta\omega(T, T_{ref})$ the variations $\Delta\omega_\varepsilon(T, T_{ref})$ and $\Delta\omega_d(T, T_{ref})$ conditioned by temperature dependences both of sapphire permittivity $\varepsilon$ and the disk dimensions are deducted by means of subtracting the corresponding curves of $f(T) = \omega/2\pi$ in Fig.1. The value of $X_s(T)$ was determined from the measured dependence $\Delta X_s(T)$, calibrated using the value $\lambda(0) = 208$nm known from the previous measurement [19].

The temperature variation of London penetration depth, $\Delta\lambda(T)$, determined from microwave data is presented in Fig.2. The observed dependence of $\Delta\lambda(T)$ follows a power law, $\Delta\lambda(T) \sim T^n$ with $n = 2.8$ from low temperatures up to at least $0.6T_c$. The obtained dependence is similar to radiofrequency range measurements [7,16,26,27], although $n$ is rather distinguished from them. When the given work results were processed, a work [28] was arrived indicatig $n \cong 2.66$. The difference $\Delta\lambda(T) = \lambda(T) - \lambda(0)$ and the superfluid density, $n_s(T) = [\lambda(0)/\lambda(T)]^2$, are commonly used to analyze penetration depth data and compare the calculations making certain assumptions regarding the superconducting gap structure [29].

The temperature dependence $[\lambda(0)/\lambda(T)]^2$ is shown in Fig.2 (see inset), where one can see the calculated curves for a power law $\Delta\lambda(T) \sim T^{2.8}$ and for exponential law $\Delta\lambda(T) \sim (\pi\Delta(0)/2kT)^{1/2} \cdot \exp(-\Delta(0)/kT)$ with $\Delta(0) = 0.75$ in $kT_c$ units, where $k$ is Boltzman coefficient. At low temperatures $\Delta(0) \cong \Delta(T)$ up to $T \cong T_c/3$. One can see that at least at low temperatures both functional laws are very close. The fact means that the temperature dependences of both London penetration depth and the superfluid density indicate evident absence of nodes of superconducting gap function and allow concluding about one of the expended s-wave symmetries of the studied pnictide.

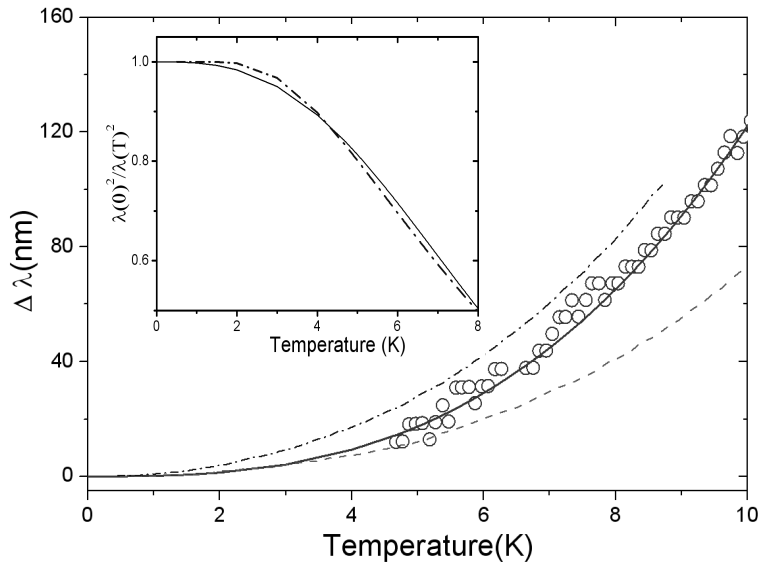

*Fig. 2.* The variarion of London penetration depth $\Delta\lambda(T)$ in a single crystal Ba(Fe$_{1-x}$Co$_x$)$_2$As$_2$ at low temperatures. The open circles represent experimental data, the solid line refers to power law $T^{2.8}$. The dushed and dot-and-dushed lines correspond to the experimental data in [7] and [27] accordingly. Inset shows the temperature dependence of superconducting electron component under condition both of the power law $\Delta\lambda(T) \sim T^{2.8}$ (solid line) and the exponential law with a small gap $\Delta/kT_c = 0.75$ (dot-and-dushed line) at low temperatures.

In summary, we carried out microwave surface impedance measurements of the optimally doped single crystal Ba(Fe$_{1-x}$Co$_x$)$_2$As$_2$ ($x = 0.074$) with critical temperature $T_c = 22.8$K and found the power-law exponent $n=2.8$ in temperature dependence of the London penetration depth. The obtained dependence is similar to radio-frequency measurements indicating no noticeable frequency dependence of the response [7]. This exponent points towards nodeless state with pairbreaking scattering, which can support one of the extended s-pairing symmetries [26]. The temperature dependence $[\lambda(0)/\lambda(T)]^2$ calculated for a power law $\Delta\lambda(T) \sim T^{2.8}$ and exponential law for one superconducting small-gap ($\Delta/kT_c = 0.75$) superconductor are very close. If another gap exsists, it has a small weight coefficient.